\documentclass[pre,aps,twocolumn,superscriptaddress,nofootinbib]{revtex4-2}
\usepackage{graphicx}
\usepackage{amssymb,latexsym,amsthm,amsmath}    
\usepackage{mathtools}
\usepackage{float} 
\usepackage[draft]{changes}
\usepackage{hyperref}
\usepackage{color}
\usepackage{hhline}
\usepackage{colortbl}
\usepackage{ulem} 
\usepackage{xcolor}

\begin{document}
\title{Cooperation transitions in social games induced by aspiration-driven players}
\author{M. Aguilar-Janita}
\email[Corresponding author:]{miguel.aguilar@urjc.es}
 \affiliation{Complex Systems Group \& GISC, Universidad Rey Juan Carlos, 28933 M\'ostoles, Spain}
 
\author{N. Khalil}%
\affiliation{Complex Systems Group \& GISC, Universidad Rey Juan Carlos, 28933 M\'ostoles, Spain}
\author{I. Leyva}
\affiliation{Complex Systems Group \& GISC, Universidad Rey Juan Carlos, 28933 M\'ostoles, Spain}
\affiliation{Center for Biomedical Technology, Universidad Polit\'ecnica de Madrid, Pozuelo de Alarc\'on, 28223 Madrid, Spain}
\author{I. Sendi\~na-Nadal}
\affiliation{Complex Systems Group \& GISC, Universidad Rey Juan Carlos, 28933 M\'ostoles, Spain}
\affiliation{Center for Biomedical Technology, Universidad Polit\'ecnica de Madrid, Pozuelo de Alarc\'on, 28223 Madrid, Spain}

\begin{abstract}
Cooperation and defection are social traits whose evolutionary origin is still unresolved. Recent behavioral experiments with humans suggested that strategy changes are driven mainly by the individuals' expectations and not by imitation. This work theoretically analyzes and numerically explores an aspiration-driven strategy updating in a well-mixed population playing games. 
The payoffs of the game matrix and the aspiration are condensed into just two parameters that allow a comprehensive description of the dynamics. We find continuous and abrupt transitions in the cooperation density with excellent agreement between theory and the Gillespie simulations. Under strong selection, the system can display several levels of steady cooperation or get trapped into absorbing states. These states are still relevant for experiments even when irrational choices are made due to their prolonged relaxation times. 
Finally, we show that for the particular case of the Prisoner Dilemma, where defection is the dominant strategy under imitation mechanisms, the self-evaluation update instead favors cooperation nonlinearly with the level of aspiration. Thus, our work provides insights into the distinct role between imitation and self-evaluation with no learning dynamics.

\end{abstract}

\maketitle

\section{Introduction}
Cooperation and defection are central to social dilemmas, where cooperative individuals contribute to the collective welfare at a personal cost, while defectors choose not to. A recent example is when someone decides to undergo physical isolation during a pandemic. This decision carries personal costs but contributes to the overall public health benefit.

Therefore, the evolutionary origin of social cooperation presents a challenging paradox due to the lower individual fitness of cooperators, but it is frequently observed in animal and human systems \cite{maynard_95,nowak_06,sigmund_10}. Evolutionary game theory provides a theoretical framework to study cooperation among individuals who interact according to predefined game rules, strategies, and payoffs. Players revise their strategies over time, leading to the emergence of an evolutionary stable strategy. 

What is the rationale behind the strategy choice has attracted much interest lately in evolutionary dynamics as it may have a significant impact on the outcomes of theoretical models and the level of cooperation in social dilemmas \cite{traulsen_pnas10}. 
There are two main mechanisms proposed for updating strategies: imitation and aspiration-driven dynamics. In imitation dynamics, individuals update their strategies by copying the more successful strategy of a peer \cite{santos_prl05,traulsen_jtb07b,wu_b_pre10,hofbauer_bams03,szabo_pr07,roca_epl09}. This spreading of successful strategies occurs through payoff-dependent learning. On the other hand, aspiration-driven updating involves individuals adjusting their strategies based on their profits compared to a value they aspire to achieve, known as the aspiration level. This update does not rely on external information about peers \cite{platkowski_aml09,macy_pnas02,posch_prslb99,perc_pone10,chen_xj_pre08,liu_x_pre16,du_j_jrsi14,du_j_sr15,lim_is_pre18}. In aspiration-driven dynamics, individuals are satisfied only if their obtained payoff exceeds a fixed aspiration level. They reinforce actions that have resulted in satisfactory outcomes and discourage those yielding unsatisfactory outcomes.

However, doubts have been raised about the actual strategy updating mechanisms. The controversy arises from experimental findings in both  two-player and  multi-player games, where it has been observed that imitation traits alone do not account for the observed strategy changes or that these changes are not solely driven by payoff comparison \cite{fischbacher_el01,grujic_pone10,grujic_pone12,grujic_srep12,gracia-lazaro_pnas12,traulsen_pnas10}. Instead, individuals respond to cooperation reciprocally, being more inclined to contribute when their partners do so, a behavior known as conditional cooperation or moody conditional cooperation if the reciprocal behavior depends on the focal player's previous action. 
These behavioral patterns have been accounted for by models incorporating dynamical updating based on aspiration learning  \cite{Cimini_JRSI14,vilone2014social,ezaki_pcb16,horita_sr17,couto_njp22}. Aspiration-driven strategy-update dynamics are commonly observed in animals  \cite{bergen_prslb04,bennett_ab08,gruter_bes11} and humans \cite{roca_pnas11} and have been considered as a single mechanism with a global fixed aspiration level \cite{du_j_jrsi14,du_j_sr15,lim_is_pre18}, heterogeneously distributed \cite{perc_pone10,zhou_l_pre18,zhou_l_nc21,wu_b_pcb18} and degree correlated aspiration distributions \cite{chen_xj_pre08}, time-dependent aspiration learning rules \cite{posch_prslb99,macy_pnas02,palomino_ijgt99,platkowski_aml09}, or combined with other updating strategy imitation-driven players \cite{liu_x_pre16,danku_epl18,wang_x_pre19,arefin_prsa21,wang_sy_amc23}.  

The role of structured interactions  in promoting cooperation has sparked debate between experiments and theory. While network reciprocity is recognized as a theoretical mechanism that promotes cooperation \cite{nowak_n92b,nakamaru_jtb97,santos_prl05,ohtsuki_jtb06,nowak_s06,szabo_pr07,perc_review2013,klemm2020altruism}, recent behavioral lab experiments with humans have shown that the level of cooperation achieved is not affected by the network topology \cite{grujic_pone10,grujic_srep12,gracia-lazaro_pnas12,gracia-lazaro_srep12}. In contrast, other experiments using artificial social networks replicating structures used in theoretical models have demonstrated that a network structure can indeed stabilize human cooperation \cite{rand_pnas14,dercole_sr19}. This suggests that specific combinations of payoffs and network structure are necessary for cooperation to succeed  \cite{Zhuk2021,khalil2023deterministic}. 

In this study, we aim to examine how a basic aspiration-driven strategy update affects the evolution of cooperation in a well-mixed population. We investigate this question in the whole spectrum of payoff possibilities for interactions between two strategies and allow for some uncertainty in the strategy adoption, which is solely determined by a constant and uniform aspiration level across the population. Despite the model's simplicity, we have identified a variety of steady and metastable states through analytical and numerical methods, depending on aspiration and game parameters and whether or not strategy choice is purely deterministic. While previous works have looked at deterministic \cite{posch_prslb99} and imperfectly rational players \cite{chen_xj_pre08,du_j_jrsi14,du_j_sr15,lim_is_pre18} with aspiration dynamics, their analyses were limited to a weak rationality regime characterized by relatively high levels of noise affecting the strategy adoption process in structured or unstructured populations, obtaining in some cases, contradictory results regarding the role of the network reciprocity. Therefore, our motivation is to further understand the baseline case by exploring analytically how the aspiration-driven rule affects the evolution of cooperation across a wider range of game, aspiration, and irrationality parameters in the mean-field case.

In Section \ref{sec:model}, we present our model where the parameters of the game are rescaled in terms of the aspiration  enabling, in Section \ref{sec:theroy}, to perform an  analytical investigation of the existence of stationary states, in good agreement with simulations. The resulting model also allows  the analysis of the critical behavior in Sec.~\ref{sec:critic},  evidencing continuous and abrupt transitions of cooperation with the aspiration and rationality levels. Finally, in Sec.~\ref{sec:PD}, we discuss the classic Prisoner Dilemma within our framework and present some conclusions in Sec.~\ref{sec:conclu}.

%%%%%%%%%%%%%%%%%%%%%%%%%%%%%%%%%%%%%%%%%%%%%%%%%%%%%%%%
\section{Model definition}
\label{sec:model}
%%%%%%%%%%%%%%%%%%%%%%%%%%%%%%%%%%%%%%%%%%%%%%%%%%%%%%%%%%

We consider a well-mixed population of \(N\) agents  having two possible strategies, cooperation (C) or defection (D). Agents play $2\times 2$ games with all others and, as a result, they receive a payoff according to the following matrix:
\begin{equation}
\begin{array}{c|cc}
 &{\rm C} & {\rm D}\\ \hhline{-|--}
{\rm C} &R  & S\\ \hhline{~|~}
{\rm D} &T & P \\
\end{array}
\end{equation}
The parameters \(R,\,S,\,T\), and \(P\) represent respectively the rewards for mutual cooperation, the sucker’s payoff, the temptation to defect, and the punishment for mutual defection. Although these parameters take restricted values in evolutionary game theory, for this study, they are considered without limitations. Hence, we do not preclude any behavior to any of the strategies. 

Provided a state of the system characterized by the cooperation fraction $\rho$, after a play round, each cooperator receives 
\begin{equation}
\label{eq:gc}
  g_c=N\left[R\left(\rho-\frac{1}{N}\right)+S(1-\rho)\right],
\end{equation}
and each defector
\begin{equation}
\label{eq:gd}
  g_d=N\left[T\rho+P\left(1-\rho-\frac{1}{N}\right)\right].
\end{equation}

These payoffs determine the system's evolution by setting the probability of a player changing her strategy from cooperation to defection ($\rho\to\rho-1/N$) or vice-versa  ($\rho\to\rho+1/N$). The corresponding  transition rates (per time unit $t_0$) are 
\begin{eqnarray}
  \label{eq:rate-}
  && \pi^-(\rho)=\frac{1}{t_0}\frac{\rho}{1+\exp\left(\frac{s_c}{\theta}\right)}, \\
  \label{eq:rate+}
  && \pi^+(\rho)=\frac{1}{t_0}\frac{1-\rho}{1+\exp\left(\frac{s_d}{\theta}\right)}, 
\end{eqnarray}
where \cite{macy_pnas02}: 
\begin{equation}
  s_r=\frac{\frac{g_r}{N-1}-m}{M_r}, \qquad r\in\{c,d\},
  \label{eq:old_s}
\end{equation}
are proportional to the difference between the current payoff of an agent and her \emph{aspiration} or \emph{mood} \(m\), which is assumed to take the same value for all agents. The payoff-aspiration difference measures the level of the agent's dissatisfaction. The factor \(\theta\) is a non-negative parameter playing the role of an effective temperature, and the normalization factors \(M_r\) are: 
\begin{eqnarray}
  && M_c=\max{(|R-m|,|S-m|)}, \\
  && M_d=\max{(|T-m|,|P-m|)}.
\end{eqnarray}

Therefore, our players follow a pure aspirational rule without social imitation nor reinforcement learning. It is essentially the model introduced and studied in \cite{posch_prslb99,du_j_jrsi14,du_j_sr15}, wherein the normalization factors \(M_r\) are taken as one, with restricted values of the game, temperature, and aspiration parameters (typically \(P=0,\, R=1\), \(m\in(0,1)\), and \(\theta\ge 0.1\)). Here, it is \(M_r\ne 1\) in general, and we let the parameters take any possible value, except for the effective temperature \(\theta\ge 0\). In particular, the aspiration parameter \(m\) can be negative, which includes social situations where agents are reluctant to change their strategy even when the received payoffs are negative (though above \(m\)). We will show that, upon removing some restrictions, the system's behavior becomes much richer than previously reported.  

It is important to notice that the model can be simplified in terms of only two parameters:
\begin{equation}
   \sigma=\frac{S-m}{|R-m|}, \qquad \tau=\frac{T-m}{|P-m|},
   \label{eq:sigma_tau}
\end{equation}
provided \(P\ne m\) and \(R\ne m\). Then, the state functions $s_r$ in Eq.~(\ref{eq:old_s}) can be written as:
\begin{eqnarray}
  && s_c=\frac{N}{N-1}\frac{\sigma(1-\rho)+k_c\left(\rho-\frac{1}{N}\right)}{\max{(1,|\sigma|)}}, \\
  && s_d=\frac{N}{N-1}\frac{\tau \rho+k_d\left(1-\rho-\frac{1}{N}\right)}{\max{(1,|\tau|)}},
\end{eqnarray}
where $k_c=\frac{R-m}{|R-m|}$ and $k_d=\frac{P-m}{|P-m|}$. We  distinguish  the following three cases:
\begin{itemize}
\item[-] Case I: $k_c$=$k_d=1$.
\item[-] Case II: $k_c$=1, $k_d=-1$.
\item[-] Case III: $k_c$=$k_d=-1$.
\end{itemize}
The  case $k_c=-1$, $k_d=1$ (case II') can be reduced to case II by interchanging the roles of cooperation and defection: \(R\leftrightarrow P\), \(\sigma\leftrightarrow \tau\), and \(\rho\leftrightarrow 1-\rho\).  

Notice that, for the typical values used in the literature \(P=0,\, R=1\), and \(m\in(0,1)\), we are always in case II, with \(k_c=-k_d=1\). Hence, all our results regarding cases I and III are completely novel. 

\section{Theory for large population}
\label{sec:theroy}

For a large, well-mixed population, the evolution of the fraction of cooperators \(\rho\) can be described by the mean-field deterministic approximation:   
\begin{equation}
  \label{eq:ecrhom}
  \frac{d\rho}{dt} =F(\rho),
\end{equation}
where
\begin{equation}
  \label{eq:effoce}
  F(\rho)=\pi^+(\rho)-\pi^-(\rho).
\end{equation}
Therefore, the fixed points $F(\rho)=0$ correspond to the steady states of the system, whose linear stability determines the reachable states. 

It is readily seen that for any \(\theta\ge 0\) the force verifies
\begin{equation}
  F(0)\ge 0,\qquad F(1)\le 0,
\end{equation}
hence, when  the effective force \(F(\rho)\) is continuous, there is at least one steady state, as we carefully analyze in the next Sections. First, we consider the case of zero effective temperature \(\theta=0\), followed by the more general situation \(\theta>0\). We will compare our predictions with numerical simulations of the master equation obtained by means of the Gillespie algorithm. Unless otherwise specified, the simulations consider a population of \(N=10^4\) agents and \(t=10^5t_0\) simulation time, while the theoretical results are  for \(N\to \infty\) and \(t=200t_0\).

%%%%%%%%%%%%%%%%%%%%%%%%%%%%%%%%%%%%%%%%%%%%%%%%%%%%%%%%%%%%%%%%%%%%%%%%%%
\subsection{Zero effective temperature}
%%%%%%%%%%%%%%%%%%%%%%%%%%%%%%%%%%%%%%%%%%%%%%%%%%%%%%%%%%%%%%%%%%%%%%%%%%

\begin{figure}[!h]
  \begin{center}
    \includegraphics[width=1\linewidth]{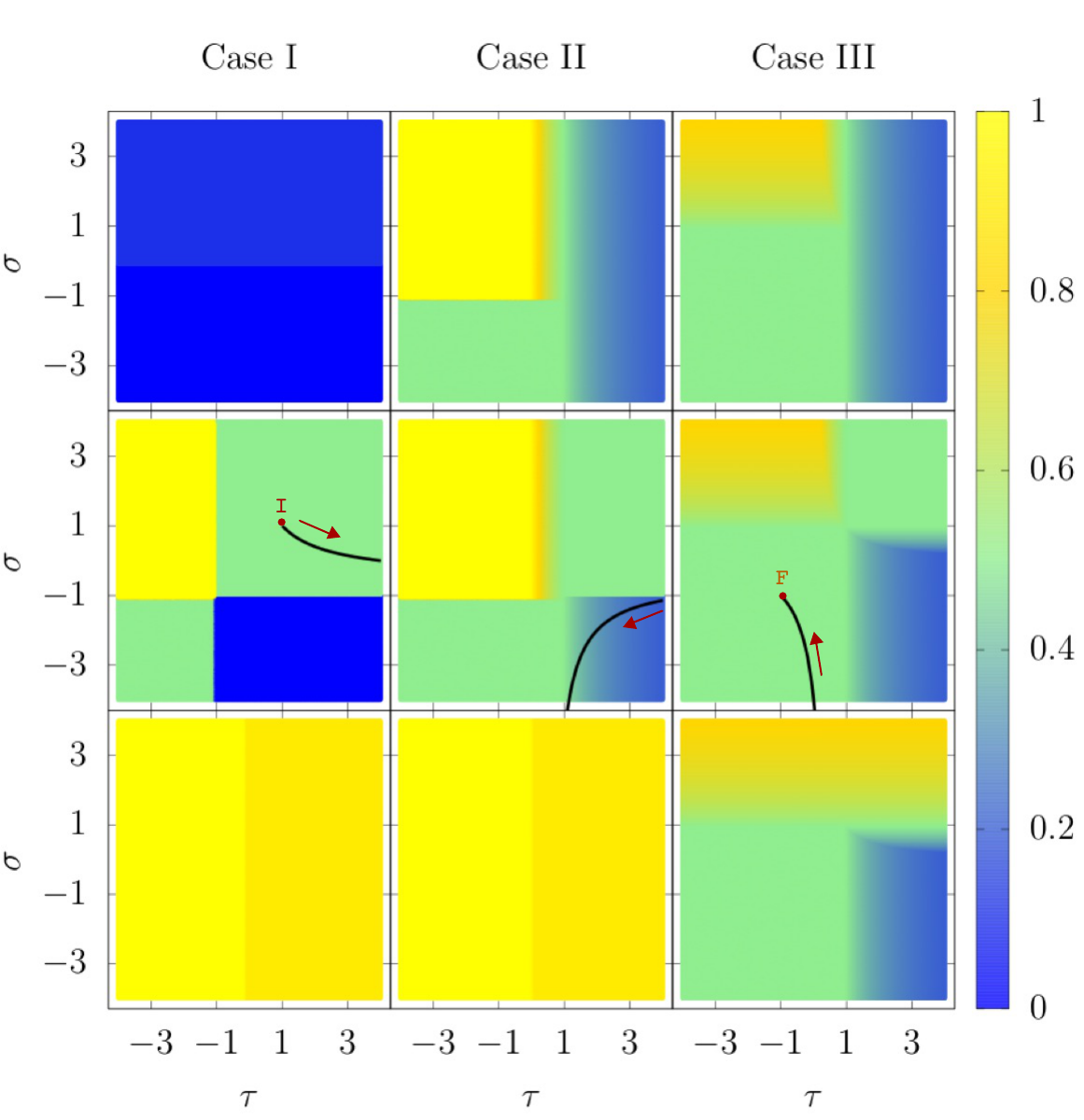} 
  \end{center}
  \caption{Numerical results for the cooperation density for zero-effective temperature \(\theta=0\). Rows correspond to different initial conditions: \(\rho_0=0.1\) (top), \(0.5\) (middle), and \(0.9\) (bottom). Columns correspond to cases I, II, and III. The black lines give (part of) the trajectory determined by fixing \(P=0,\, R=1,\, S=-\frac{1}{2}\), and \(T=\frac{3}{2}\) and increasing \(m\), from \(m=-\infty\) at point \(I=(1,1)\) (first column) to \(m=\infty\) at point \(F=(-1,-1)\) (third column), following the arrows. \label{fig:diagramatheta0} }
\end{figure}

When \(\theta=0\), \(F(\rho)\) is a piece-wise function of \(\rho\), since the transition rates in Eqs.~(\ref{eq:rate-}-\ref{eq:rate+}) take the form:
\begin{eqnarray}
  \label{eq:pimenos}
  && \pi^-(\rho)=\frac{\rho}{t_0}\Theta\left[-\sigma(1-\rho)-k_c\left(\rho-\frac{1}{N}\right)\right], \\
  \label{eq:pimas}
  && \pi^+(\rho)=\frac{1-\rho}{t_0}\Theta\left[-\tau\rho-k_d\left(1-\rho-\frac{1}{N}\right)\right], 
\end{eqnarray}
where \(\Theta\) is the Heaviside step function [we take \(\Theta(0)=1/2\)]. We summarize the numerical results in Fig.~\ref{fig:diagramatheta0}, where the final state of the cooperation fraction \(\rho\) is represented in the \(\sigma-\tau\) space for the three previously distinguished cases depending on the values of the pair  $(k_c,k_d)$, and for several initial conditions.   

% FIGURE 1 %%%%%%%%%%%%%%%%%%%%%%%%%%%%%%%%%%%%%%%%%%%%%%%%%%%%%%%%%
%%%%%%%%%%%%%%%%%%%%%%%%%%%%%%%%%%%%%%%%%%%%%%%%%%%%%%%%%%%%%%%%%%%%%%%%%%

%%%%%%%%FIGURE 2 %%%%%%%%%%%%%%%%%%%%%%%%%%%%%%%%%%%%%%%%%%%%%%%%

\begin{figure*}[!t]
  \begin{center}
    \includegraphics[width=0.99\linewidth]{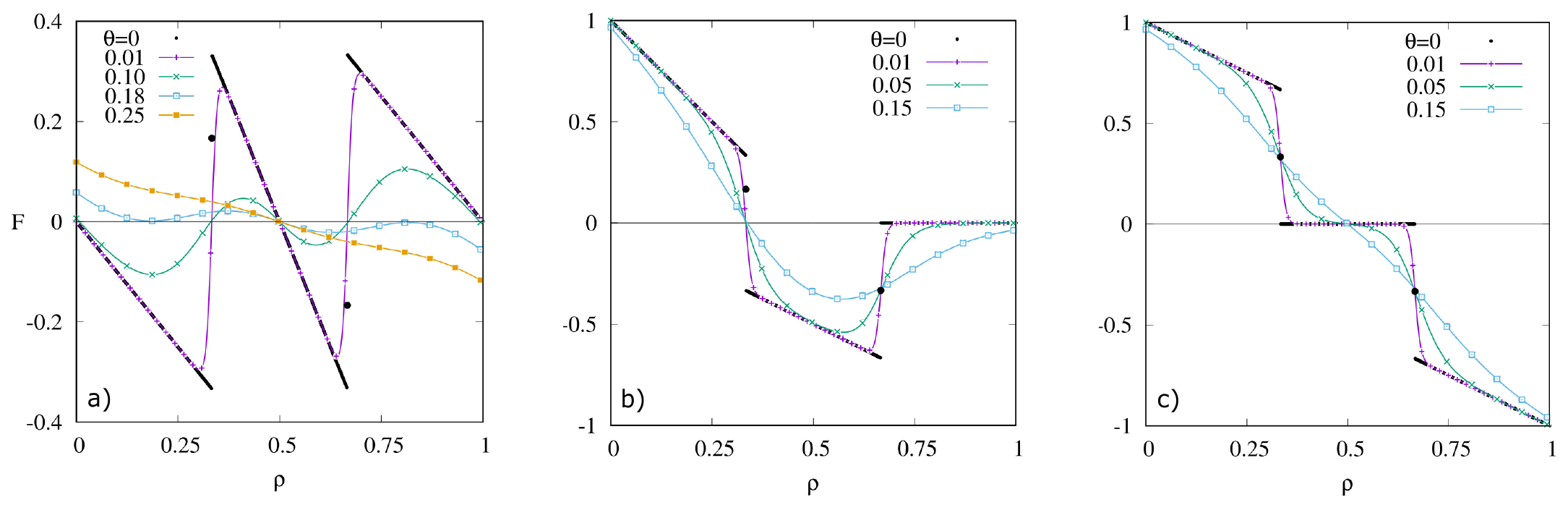}  
  \end{center}
  \caption{Force function \(F(\rho)\) given by Eq.~\eqref{eq:effoce} for several effective temperatures and (a) \(\tau=\sigma=-2\) of case I (\(k_c=k_d=1\)), (b) \(\tau=-\sigma=2\) of case II (\(k_c=-k_d=1\)) , and (c) \(\tau=\sigma=2\) of case III (\(k_c=k_d=-1\)). In all panels, solid lines correspond to theory while symbols correspond to numerical simulations.\label{fig:abcForce}}
\end{figure*}
%%%%%%%%%%%%%%%%%%%%%%%%%%%%%%%%%%%%%%%%%%%%%%%%%%%%%%%%%%%%%%%%%%%%%%%%%%

In our research, we have identified several fixed points - including consensus, coexistence, absorbing, and discontinuous states - that possess interesting properties. We have confirmed this through numerical simulations, which we will discuss in the following paragraphs. This analysis of the steady states of $F(\rho)=0$, will be further extended in Appendix \ref{appen:cmedio}. 

%%%%%%%%%%%%%%%%%%%%%%%%%%%%%%%%%%%%%%%%%%%%%%%%%%%%%%%%%%%%%%%%%%%%%%%%%%
\subsubsection{Consensus states}

Consensus refers to situations where everyone either cooperates or defects. Full cooperation \(\rho=1\) is only linearly stable when $k_c=1$ and \(\tau<0\) (cases I and II). Similarly, full defection \(\rho=0\) is only linearly stable when $k_d=1$ and \(\sigma<0\) (cases I and II'). It is worth noting that the solution for case II'  can be obtained from case II using the procedure described in the previous section. 

This theoretical analysis agrees with the numerical results shown in Fig.~\ref{fig:diagramatheta0}, as long as the initial conditions $\rho_0$ are appropriate.  This last point is better understood in the example of Figure~\ref{fig:abcForce}(a) where the solid black line depicts \(F(\rho)\) vs $\rho$ for $\theta=0$ and case I.  We see that for any initial cooperation fraction $\rho_0<\frac{1}{3}$ the system tends to full defection \(\rho=0\), while for  $\rho_0>\frac{2}{3}$ full cooperation is reached. 
 
%%%%%%%%%%%%%%%%%%%%%%%%%%%%%%%%%%%%%%%%%%%%%%%%%%%%%%%%%%%%%%%%%%%%%%%%%%
\subsubsection{Coexistence states}
 
By coexistence, we refer to a state where \(\rho=\frac{1}{2}\). By imposing the steady-state condition, \(\pi^-(\tfrac{1}{2})=\pi^+(\tfrac{1}{2})\), we  conclude that coexistence is the only potential steady-state solution when both strategies have payoffs larger or equal than the aspiration or mood $m$.
For the latter to occur, we need 
\begin{eqnarray}
  \label{eq:coex1}
  && \sigma \le -\frac{N-2}{N}k_c\simeq -k_c, \\
  \label{eq:coex2}
  && \tau \le -\frac{N-2}{N}k_d\simeq -k_d.
\end{eqnarray}
Assuming that the arguments of the \(\Theta\) functions are positive, the coexistence solution is the sole linearly stable option in these circumstances.  Figure \ref{fig:abcForce}(a) (black line) shows that coexistence occurs when the initial condition falls within the range of $\rho_0\in \left[\frac{1}{3},\frac{2}{3}\right]$. Otherwise, the system will reach a consensus. This is a clear sign of a discontinuous transition in the final cooperation level depending on the initial condition, which will be further examined later.

The middle panel of Fig.~\ref{fig:diagramatheta0} shows the  numerical solution for the coexistence states when $\rho_0=\frac{1}{2}$. Although the solution $\rho=\frac{1}{2}$ may also appear for other parameter values that do not satisfy Eqs.~\eqref{eq:coex1} and \eqref{eq:coex2}, as seen in Fig.~\ref{fig:diagramatheta0}, the nature of those solutions is different and  will be classified as absorbing states  in the next section. 

%%%%%%%%%%%%%%%%%%%%%%%%%%%%%%%%%%%%%%%%%%%%%%%%
\subsubsection{Absorbing states}
 
All values of \(\rho\) that make the arguments of both \(\Theta\) functions in Eqs.~\eqref{eq:pimenos} and \eqref{eq:pimas} negative are marginally stable solutions of Eq.~\eqref{eq:ecrhom}. The conditions read:
\begin{eqnarray}
  \label{eq:absorb1}
  && \sigma(1-\rho)+k_c\left(\rho-\frac{1}{N}\right)>0, \\
  \label{eq:absorb2}
  && \tau\rho+k_d\left(1-\rho-\frac{1}{N}\right)>0.
\end{eqnarray}
These solutions are absorbing states: the system becomes stuck once it reaches any of them (either from another non-absorbing state or due to the initial conditions). This occurs because all probability transitions are zero, which is valid beyond the infinite-system approximation but only within the zero-temperature case. It is important to note that, strictly speaking, the absorbing states disappear for $\theta>0$ but, instead, the system may reach a metastable state (with a large relaxation time), as we show later.  

We present Fig.~\ref{fig:abcForce}(c) (black line) as an instance of a force \(F(\rho)\) where the only steady states are absorbing, and depending on the initial value \(\rho_0\), the system will approach to or remain in the interval $[\frac13,\frac23]$. In this case, the coexistence state \(\rho=\frac12\) is one of the absorbing states. We show another example in Fig.~\ref{fig:abcForce}(b), where the absorbing states can only be attained from the initial state when \(\rho_0\in[\frac{2}{3},1]\). The consensus state \(\rho=1\) is also absorbing in this latter case.  

%%%%%%%%%%%%%%%%%%%%%%%%%%%%%%%%%%%%%%%%%%%%%%%%%%%%%%%%%%%%%%%%%%%%%%%%%%
\subsubsection{Discontinuous states}

Finally, the system can also reach another steady state, which we call discontinuous due to its (eventual) appearance at a discontinuity of \(F(\rho)\). To describe the new solutions, take as an example the function force \(F(\rho)\) depicted in Fig.~\ref{fig:abcForce}(b) (black line) where the discontinuities are at \(\rho=\frac13\) and \(\frac23\). If initially \(\rho\in[0,\frac{1}{3}]\) or \(\rho\in[\frac{1}{3},\frac{2}{3}]\), the cooperation fraction tends to \(\rho^*=\frac13\). This is a value for which \(F(\rho^*)\ne 0\) but is still  an attractor of the dynamics. The other discontinuity at \(\rho=\frac{2}{3}\) does not have this property. In general, in order for \(\rho^*\) to represent a discontinuous state, we need \(F(\rho^{*-})>0\) and \(F(\rho^{*+})<0\). Upon slightly increasing the effective temperature, the discontinuous states become (conventional) stable fixed points, as discussed below.

%%%%%%%%%%%%%%%%%%%%%%%%% FIGURE 3 %%%%%%%%%%%%%%%%%%%%%%%%%%%%%%%%%%%%%%%%%%%%%%
\begin{figure*}[!t]
  \begin{center}
      \includegraphics[width=1\textwidth]{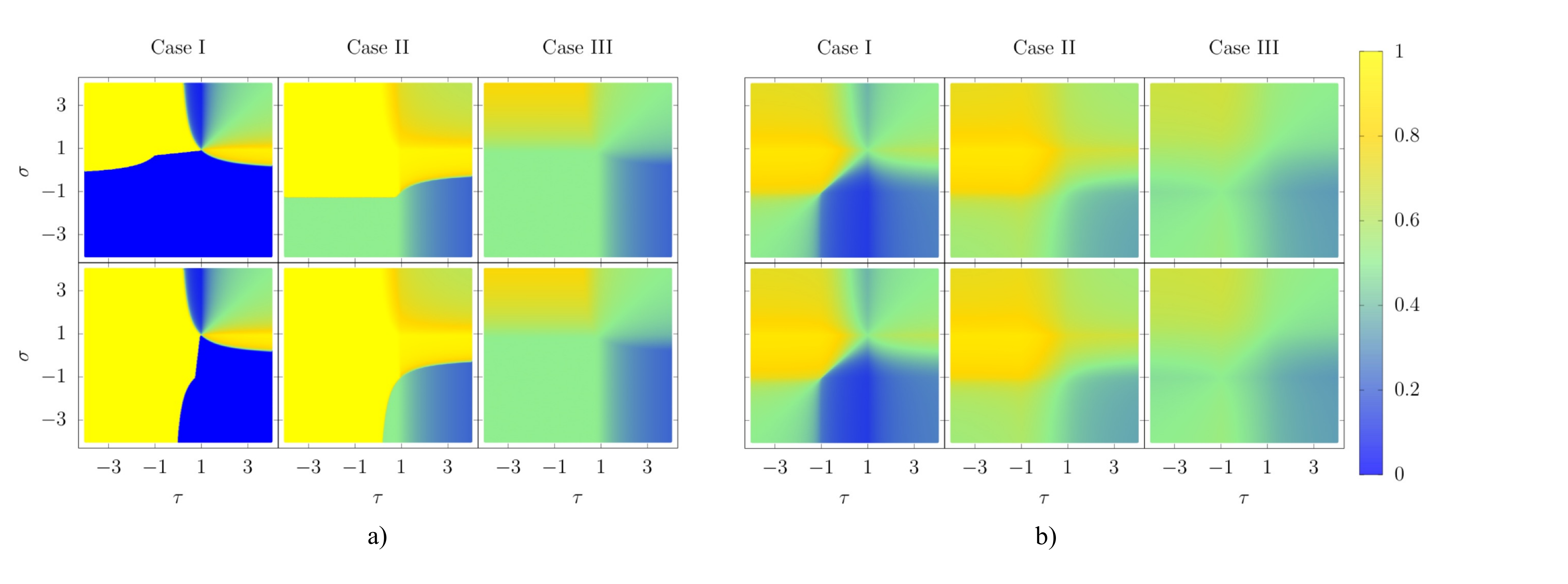}
  \end{center}
  \caption{Simulation results for the cooperation density as a function of $\tau$ and $\sigma$ for an effective temperature (a) $\theta=0.02$ and (b) $\theta=0.5$. Top row: initial condition \(\rho_0=0.1\), bottom row: \(\rho_0=0.9\). \label{fig:diagramatheta}}
\end{figure*}
%%%%%%%%%%%%%%%%%%%%%%%%% FIGURE 3 %%%%%%%%%%%%%%%%%%%%%%%%%%%%%%%%%%%%%%%%%%%%%%

%%%%%%%%%%%%%%%%%%%%%%%%%%%%%%%%%%%%%%%%%%%%%%%%%%%%%%%%%%%%%%%%%%%%%%%
\subsection{Positive effective temperature}
%%%%%%%%%%%%%%%%%%%%%%%%%%%%%%%%%%%%%%%%%%%%%%%%%%%%%%%%%%%%%%%%%%%%%%%

We now consider positive effective temperature \(\theta>0\). In this case, the cooperation fraction \(\rho\) still obeys Eq.~\eqref{eq:ecrhom} but now the rates are approximately (\(N\gg 1\)) given by 
\begin{eqnarray}
  && \pi^-(\rho)\simeq \frac{1}{t_0}\frac{\rho}{1+\exp\left[\frac{\sigma(1-\rho)+k_c\rho}{\theta \max{(1,|\sigma|)}}\right]},\\
  && \pi^+(\rho)\simeq \frac{1}{t_0}\frac{1-\rho}{1+\exp\left[\frac{\tau \rho+k_d(1-\rho)}{\theta \max{(1,|\tau|)}}\right]}.
\end{eqnarray}

The temperature increase has a profound effect on the dynamics and steady states of the system. First,  for \(\theta>0\) the force \(F(\rho)\) becomes a continuous function of \(\rho\), which means that the discontinuous states disappear. Moreover, there are no absorbing states since both rates \(\pi^\pm\) are positive (non-zero) functions for all \(\rho\). Second, the force verifies $F(0)>0$, $F(1)<0$,
that is, the consensus states also disappear. Therefore, only the coexistence state \(\rho=\frac12\) may survive for \(\theta>0\).
Nevertheless, new isolated steady states eventually emerge, with the same mathematical properties as the coexistence state but with possibly different cooperation fractions. Let us now reconsider the specific examples given in Fig.~\ref{fig:abcForce}.

We first focus on case I in Fig.~\ref{fig:abcForce}(a). For weak selection intensity \(\theta>0\), the consensus states move to two nearby stable states. Additionally, due to the continuity of $F(\rho)$, two unstable states arise near \(\rho=\frac{1}{3}\) and \(\frac{2}{3}\).  Upon further increasing  of \(\theta\), the former states approach $\rho=\frac12$ until they suddenly disappear at \(\theta\simeq 0.18\). This is an instance of a discontinuous transition induced by the effective temperature.  Finally, in the limit of high temperature, only the coexistence state survives since \(F(\rho)= \frac{1}{2t_0} (1-2\rho)\). As far as we know, the eventual appearance of more than one final state as the rationality of the agents in choosing strategy diminishes has not been previously reported in models of aspiration-driven dynamics since usually only the regime of large temperature or weak-selection limit has been considered so far. 

For case III shown in Fig.~\ref{fig:abcForce}(c), a positive temperature causes the absorbing states to disappear and reduces it to coexistence. In cases where \(\theta\) is sufficiently small but positive, the system's relaxation process can be very slow, resulting in states that are absorbing in practice. Finally, for case II [Fig.~\ref{fig:abcForce}(b)], we see an example where the interval of absorbing states for \(\rho>\frac{1}{6}\) becomes a repulsive region for $\theta>0$. Looking at the same figure for \(\rho\simeq \frac{1}{3}\), we conclude that the discontinuous states become stable steady states when \(\theta\) increases.

In Fig.~\ref{fig:diagramatheta}, we can see the final system's state in the $(\tau,\sigma)$ plane for two initial conditions, $\rho_0=0.1$ (top row) and $0.9$ (bottom row), and for two temperatures, a mild one \(\theta=0.02\) [panels a)],  and a moderate one \(\theta=0.5\) [panels b)]. It is very informative to compare this figure with the deterministic case shown in Fig.~\ref{fig:diagramatheta0} for \(\theta=0\). When the temperature is low (\(\theta=0.02\)), the diagram shows abrupt transitions but with a different topology than  \(\theta=0\). There are large regions where the final state still depends on the initial conditions. However, when the temperature is moderate (\(\theta=0.5\)), the final state is independent of the initial cooperation fraction and smoothly depends on \(\tau,\,\sigma\) in most regions. In this case, regions are close to coexistence, which is the only possible state when \(\theta\to \infty\).

%%%%%%%%%%%%%%%%%% FIGURe %%%%%%%%%%%%%%%%%%%%%%%%%%%%%%%%%%%%%5
\begin{figure*}[!t]
  \begin{center}
    \includegraphics[width=.99\linewidth]{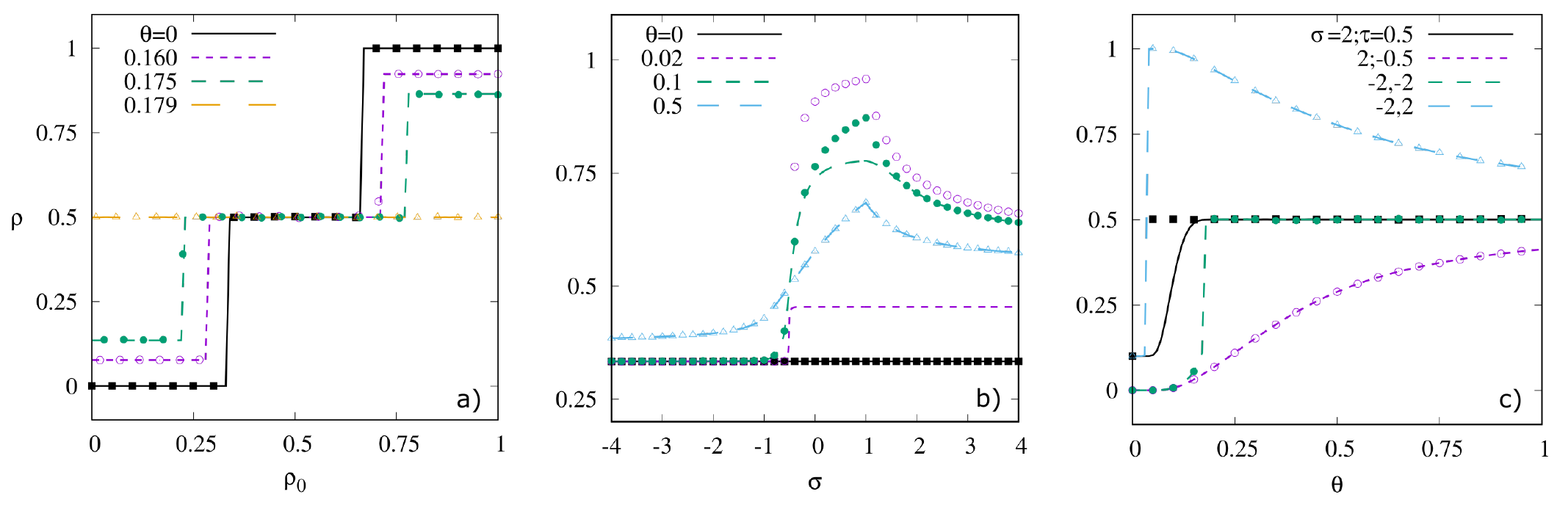}
  \end{center}
  \caption{Cooperation density \(\rho\) at \(t=200t_0\) (theory-lines) and \(t\sim 10^5t_0\) (simulations-symbols) as a function of (a) the initial value \(\rho_0\) for different \(\theta\)'s, (b) the game parameter \(\sigma\) for different \(\theta\)'s and \(\rho_0=0.1\), and (c) the effective temperature for different game parameters and \(\rho_0=0.1\). \label{fig:multi_ci} }
\end{figure*}
%%%%%%%%%%%%%%%%%% FIGURe %%%%%%%%%%%%%%%%%%%%%%%%%%%%%%%%%%%%%5

%%%%%%%%%%%%%%%%%%%%%%%%%%%%%%%%%%%%%%%%%%%%%%%%%%%%%%%%%%%%%%%%%%%%%%%%%%
\section{Critical behaviour}\label{sec:critic}
%%%%%%%%%%%%%%%%%%%%%%%%%%%%%%%%%%%%%%%%%%%%%%%%%%%%%%%%%%%%%%%%%%%%%%%%%%
As shown in the previous section, this reduced game scenario where the player's strategy is determined solely by their inner aspirations still presents a complex situation with multiple critical transitions. Here, we complete the picture by paying close attention to the transitions as a function of the main parameters of the system. Again, we  validate our analytical predictions for well-mixed, $N\to\infty$ with simulations for large populations.  

%%%%%%%%%%%%%%%%%%%%%%%%%%%%%%%%%%%%%%%%%%%%%%%%%%%%%%%%%%%%%%%%%%%%%%%%%
%\subsection{Initial condition} 

We first examine the relationship between the initial condition $\rho_0$ and the final cooperation fraction \(\rho\) in Fig.~\ref{fig:multi_ci}(a). We utilize the parameter values depicted in Fig.~\ref{fig:abcForce} whenever applicable. We observe that, at low noise levels (low irrationality), the sharp transitions seen for $\theta$=0 are maintained. The dependence of the final state on the initial conditions was already recognized in previous works for \(\theta=0\), see, for instance, Ref. \cite{posch_prslb99}. Here we show that this is also the case for small enough temperatures and some values of the game and aspiration parameters. At moderate levels, the system becomes indifferent to $\rho_0$ and achieves a state near coexistence, irrespective of the initial conditions. This behavior is consistent with the known results in the so-called weak-selection limit \cite{du_j_jrsi14,du_j_sr15,lim_is_pre18}.

The system's abrupt transitions, when analyzed as a function of the game parameter $\sigma$ in Fig.~\ref{fig:multi_ci}(b), are particularly interesting. These transitions persist through moderate levels of noise (moderate irrationality), as previously observed in Fig.~\ref{fig:diagramatheta}. However, there is a discrepancy between theory and simulations for weak noise due to the long transient times in regimes where $F(\rho)$ is small, which results in a relaxation period longer than the used one, \(200t_0\). However, this effect disappears in the moderate and strong noise regime.

%%%%%%%%%%%%%%%%%% FIGURe %%%%%%%%%%%%%%%%%%%%%%%%%%%%%%%%%%%%%5
%\begin{figure}[!h]
%  \begin{center}
%    \includegraphics[width=1\linewidth]{figs_article/multi_param.pdf}
%  \end{center}
%  \caption{Cooperation density \(\rho\) at \(t=200t_0\) (theory) and \(t\sim 10^5\) (simulations) for the initial value \(\rho_0=0.1\) and game parameter \(\tau=2\). Left: \(k_d=k_c=1\); center:  \(k_d=k_c=-1\); right: \(k_d=-k_c=-1\). Lines are for theory and symbols for numerical simulations. \il{ Creo que solo pondría uno de los panels como illustration, son cualitativamente bastante parecidos. No se distinguen bien las lineas de los puntos, sobre todo los morados y verdes. Si se aumentara el tiempo de simulacion mejoraria significativamente el parecido teoria/simulaciones? }
%  \label{fig:multi_param}}
%\end{figure}
%%%%%%%%%%%%%%%%%% FIGURe %%%%%%%%%%%%%%%%%%%%%%%%%%%%%%%%%%%%%5

\begin{figure}[!h]
  \begin{center}
    \includegraphics[width=.75\linewidth]{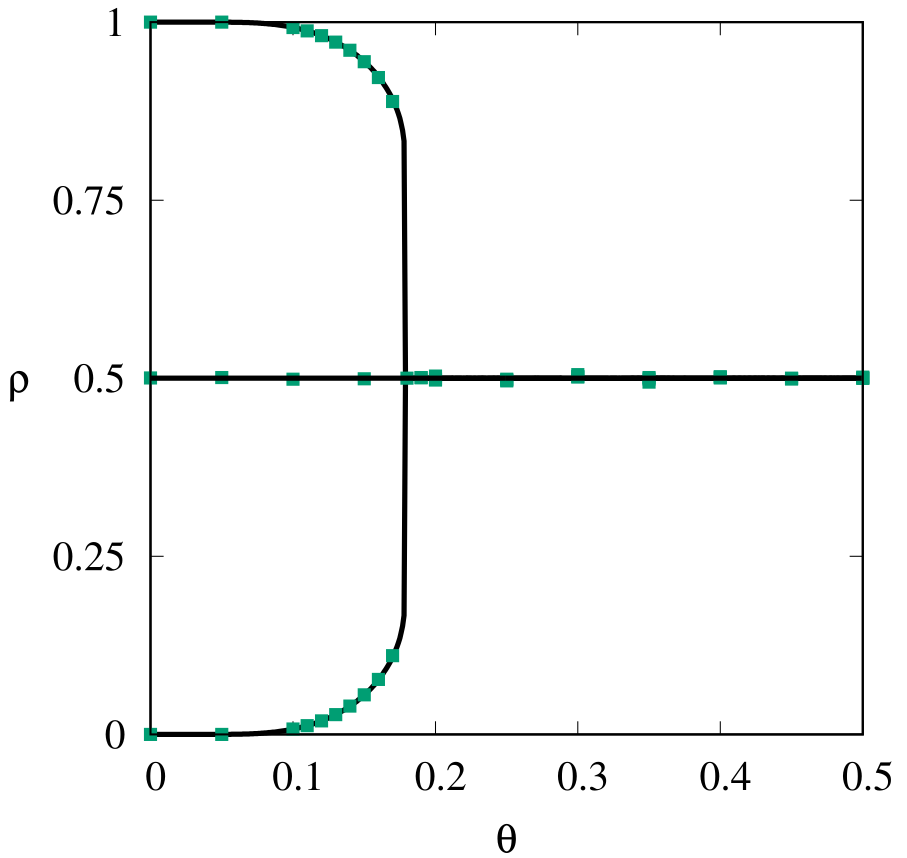}
  \end{center}
  \caption{Steady cooperation level \(\rho\) for different initial conditions as a function of the effective temperature \(\theta\), with the other parameters taken as in Fig.~\ref{fig:abcForce}(a). Black lines are from the mean-field theory and symbols from numerical simulations. \label{fig:saddle}}
\end{figure}

Finally, we have previously provided examples in which, for a large part of the parameter space, the effect of increasing \(\theta\) is pushing the system towards coexistence. However,  we show in Fig.,~\ref{fig:multi_ci}(c) an example of how weak noise can also have a constructing effect, inducing new abrupt transitions and/or non-monotonous approaches to coexistence.  

As another example of a temperature-induced transition, in Fig.~\ref{fig:saddle} we show how the cooperation density \(\rho\) passes through two saddle-node bifurcations as temperature rises, depending on the initial conditions. Namely, for temperature \(\theta\) small enough, and for the appropriate initial conditions, the system can be in a state close to defection (left zero of the force function \(F\) in Fig.~\ref{fig:abcForce}(a), close to cooperation (right zero of \(F\)), or at coexistence (central zero of \(F\)). As \(\theta\) increases, the former states continuously approach \(\rho=\frac{1}{2}\) until they suddenly disappear (the extreme stable fixed points of \(F\) collide with unstable ones). Just after these critical points, the cooperation level jumps to coexistence \(\rho=\frac{1}{2}\). We have just described two instances of a hybrid phase transition, a continuous one just after the critical point and a discontinuous one just before it. Similar behaviors have been observed in many other complex systems \cite{gomez2011explosive,baxter2015critical,cai2015avalanche,lee2016hybrid}. Interestingly, the critical behavior is found here at the mean-field level, without the need for any underlying complex interaction structure. 

%%%%%%%%%%%%%%%%%%%%%%%%%%%%%%%%%%%%%%%%%%%%%%%%%%%%%%%%%%%%%%%%%%%%%%%%%%%%%%
\section{The Prisoner's dilemma}\label{sec:PD}
%%%%%%%%%%%%%%%%%%%%%%%%%%%%%%%%%%%%%%%%%%%%%%%%%%%%%%%%%%%%%%%%%%%%%%%%%%%%%%%

Up to this point, we have demonstrated the usefulness of the scaled parameters \(\tau\) and \(\sigma\) in simplifying the description of the system. Nonetheless, to make comparisons with prior findings \cite{khalil2023deterministic} and others, in this Section we will delve into the system dynamics in terms of its intrinsic parameters, including the aspiration $m$. We also take representative game parameters corresponding to the Prisoner's dilemma game: \(P=0,\, R=1,\, S=-\frac{1}{2}\), and \(T=\frac{3}{2}\) (results for other games can be found in Appendix B).  

\begin{figure}[!h]
  \begin{center}
    \includegraphics[width=0.99\linewidth]{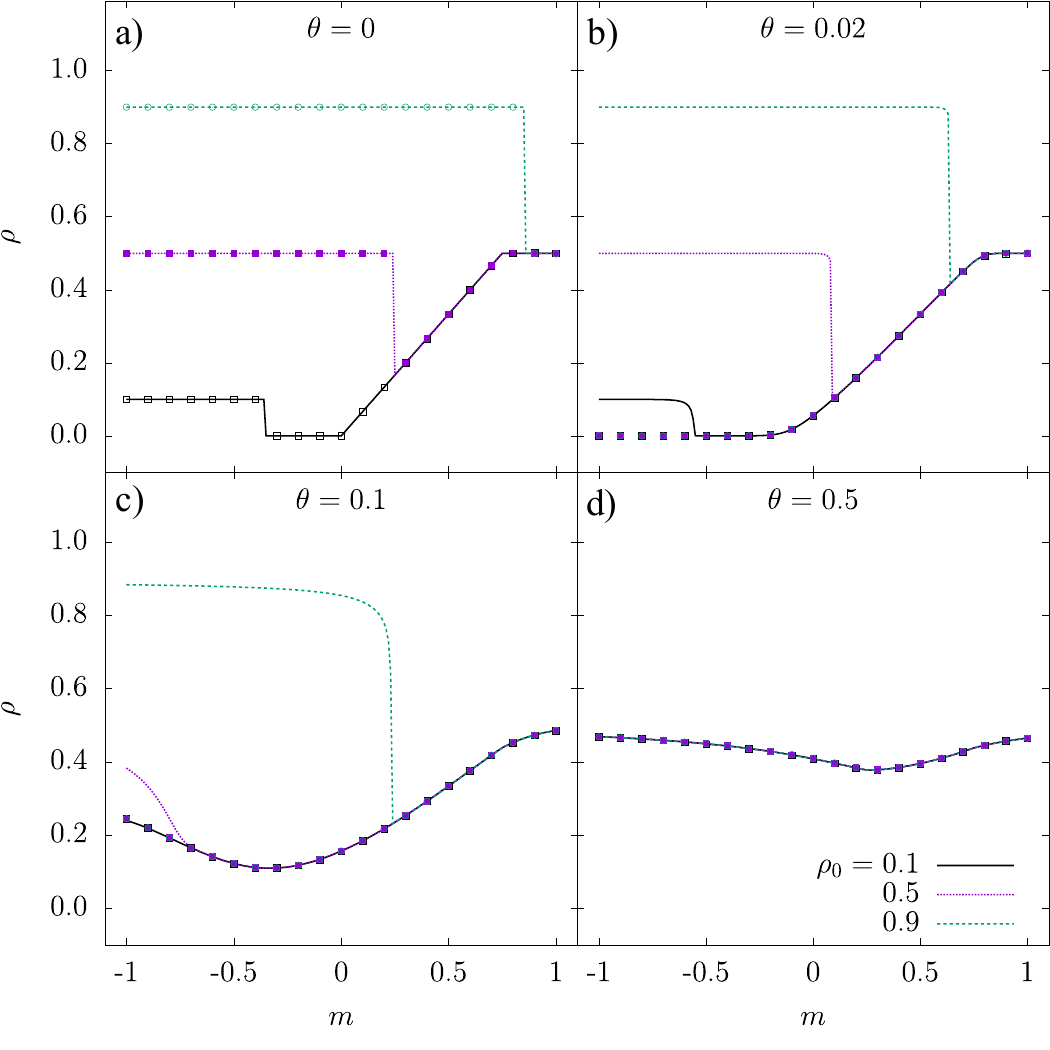}
  \end{center}
  \caption{Cooperation density \(\rho\) as a function of the mood \(m\) for three initial values \(\rho_0=0.1;0.5;0.9\), the game parameters \(P=0;\, R=1;\, S=-0.5;\, T=1.5\), and (a) \(\theta=0\), (b) \(\theta=0.02\), (c) \(\theta=0.1\), and (d) \(\theta=0.5\). Lines are for theory (\(t=200t_0\)) and symbols for numerical simulations (\(t\sim 10^5t_0\)). \label{fig:games1}}
\end{figure}

In the case of imitative strategies, a well-known result is that these parameters would yield full defection in structureless populations unless the level of noise (irrationality) is high. In the present model, we illustrate the dependence on $m$ in Fig.~\ref{fig:games1}. In this figure, we provide the final state \(\rho\) after a transient time (\(200t_0\) for theory and \(\sim10^5t_0\) in simulations), for several values of the initial condition $\rho_0$ and effective temperature $\theta$. 

We see in Fig. \ref{fig:games1} that, in general, the system does not decay to total  defection.
In the deterministic case \(\theta=0\) [Fig.~\ref{fig:games1}(a)], for small $m$ any state of the system is absorbing because any payoff is above the threshold required for strategy changes.  This cooperation level suddenly drops at some $m_c(\rho_0)$, where cooperators become unsatisfied with their payoffs and change their strategies to defection. Finally, for very large $m$ all agents change their strategy at every time step, resulting in coexistence. 

%%%%%%%%%%%%%%%%%%%%%%FIGURE%%%%%%%%%%%%%%%%%%%%%%%%%%%%%%%%
%%%%%%%%%%%%%%%%%%%%%%FIGURE%%%%%%%%%%%%%%%%%%%%%%%%%%%%%%%%

For larger values of \(m\), larger than \(\sim 0.8\) in Fig.~\ref{fig:games1}(a), the system ends up having \(\rho=\frac{1}{2}\). This cooperation value has been identified in the mean-field description as coexistence (or eventually as absorbing when \(\theta=0\)). However, \(\rho=\frac12\) does not necessarily mean that the same fraction of the population has different strategies at a given time. What really happens at the microscopic level is that all agents are dissatisfied with their payoffs, making them change their strategies each game round. In our microscopic implementation of the model by means of the Gillespie algorithm, which uses asynchronous updates, the system does reach pure coexistence of opinions (half of the population coexisting with different opinions at a given time). However, with synchronous updates, the system would change its initial  cooperation density \(\rho_0\) to \(1-\rho_0\) after a game round, and then back to \(\rho_0\), and so on. Upon time averaging, the density fraction takes the value \(\frac12\). The mean-field description of both implementations is identical and the two microscopic scenarios lead to the same macroscopic state. 

For small values of noise [Fig.~\ref{fig:games1}(b)], we notice that the observed state of the system can be strongly dependent on the evolution time, which can be very relevant in experiments. For times of the order of \(200t_0\) (comparable to typical experimental time in social experiments), the results are similar to $\theta=0$, while for longer times the system escapes to a lower cooperation level, which turns out to be independent of $\rho_0$. It is worth noting that as the aspiration (\(m\)) increases, the cooperation density (\(\rho\)) also increases.  Therefore, we conclude that aspiration promotes cooperation in Prisoner's dilemma games dominated by mainly rational choices.  
For higher noise levels, the system evolves faster to a coexistence level that shows a non-monotonous dependence on the aspiration $m$ [Figs.~\ref{fig:games1}(c-d)]. 

%%%%%%%%%%%%%%%%%%%%%%%%%%%%%%%%%%%%%%%%%%%%%%%%%%%%%%%%%%%%%
\section{Conclusions}\label{sec:conclu}
%%%%%%%%%%%%%%%%%%%%%%%%%%%%%%%%%%%%%%%%%%%%%%%%%%%%%%%%%%%%%%

The behavior of introspective agents in the context of mean-field evolutionary games has been studied theoretically and numerically. In the considered model, the strategic choices are based only on the weighting of their obtained payoff with her aspiration. An effective temperature has also been introduced to tune the degree of rationality in decision-making. Overall, the results show a rich phenomenology, which includes continuous and abrupt transitions and non-monotonous dependence of the cooperation density as the parameters of the game, effective temperature, and aspiration change. Interestingly, the general tendency is that the dynamics promote cooperation as the aspiration rises. 

We have shown that the mean-field description, in terms of the cooperation density, can be made using the effective temperature and two global parameters, a combination of the game and aspiration parameters. This means that a change in the aspiration can be seen as a change in the game parameters. Moreover, we have shown that the dynamics mimics that of a 1D overdamped particles subject to an effective force, Eq.~\eqref{eq:effoce}, which encloses all the relevant information to determine the steady-state and metastable states of the system. The analysis of the external force reveals different steady-states (consensus, coexistence, absorbing, and discontinuous) for rational decisions (zero effective temperature) and states with an intermediate degree of cooperation (similar to coexistence) with some degree of irrationality (positive temperature). The rich behavior of the system is also due to its dependence on the initial conditions, since at given values of the parameters, several steady-states may compete, as summarized in Figs. \ref{fig:diagramatheta0} and \ref{fig:diagramatheta}. 

Our study emphasizes the importance of the absorbing states when the dynamics is driven by rational decisions. Absorbing states, where the system gets stacked, are reached when players are satisfied with their payoffs (states of satisfaction) and hold the same strategy all time. We stress that these states appear at the mean-field level, i.e. without the need for any complex interaction structure, as opposed to other models \cite{khalil2023deterministic}. With some degree of irrationality, the system may effectively get trapped in (metastable) states of satisfaction, since the time needed for the system to change its state is typically much larger than the relevant ones in Sociology. This is in close relation to the eventual dependence on the system size, which will be explored elsewhere. 

The results of the present work stressed that, in order to have a complete picture of evolutionary game theory, in general, and with  aspiration-driven models, in particular, one has to be careful with the restriction on parameter values. In our case, we have identified two relevant parameters that give rise to four possible cases where the system has different behaviors. With the typical values of the game parameters \(P=0\) and \(R=1\) and with the aspiration parameter \(m\in(0,1)\), only one of the four possible cases can be covered. This leaves out relevant results.

\section*{Acknowledgments}
This research was supported by the Spanish  Ministerio de Ciencia e Innovación (Project PID2020-113737GB-I00) and by Community of Madrid and Rey Juan Carlos University through Young Researchers program in R\&D (Grant CCASSE M2737).  

%%%%%%%%%%%%%%%%%%%%%%%%%%%%%%%%%%%%%%%%%%%%%%%%%%%%%%%%%%%%%%%%%%%%%%%%%%%%%%
\appendix
%%%%%%%%%%%%%%%%%%%%%%%%%%%%%%%%%%%%%%%%%%%%%%%%%%%%%%%%%%%%%%%%%%%%%%%%%%%%%%

\section{Further results for \(\theta=0\)}
\label{appen:cmedio}
Here we complement the results of Sec.~\ref{sec:theroy} by providing the values of the parameters where the different steady states may appear for the zero-temperature case \(\theta=0\). The fundamental equation is
\begin{equation}
  \label{eq:force0}
  F(\rho)=0,
\end{equation}
where the effective force \(F\) is given by Eq.~\eqref{eq:effoce} and Eqs.\eqref{eq:pimenos}--\eqref{eq:pimas}. 

\subsection{Consensus states}

Plugging \(\rho=1\) into Eq.~\eqref{eq:force0} we obtain the condition
\begin{equation}
  \pi^-(1)=0 \Rightarrow \Theta\left(-k_c\frac{N-1}{N}\right)=0,
\end{equation}
which is verified if and only if \(k_c>0\), i.e. when \(R>m\). Disregarding border cases [i.e. assuming \(\pi^-(\rho)=0\) for \(\rho\sim 1\)], the solution \(\rho=1\), provided it exists, can be linearly stable or marginal. In the former case \(T<m \Rightarrow \pi^+=\frac{1-\rho}{t_0}\) and \(\frac{d}{dt}\rho=\frac{1-\rho}{t_0}\) for \(\rho\sim 1\) which has a stable fixed point at \(\rho=1\). In the other case, \(\frac{d}{dt}\rho=0\) for \(\rho\sim 1\) and the stability of \(\rho=1\) is marginal.

Proceeding analogously with the defection consensus \(\rho=0\), we find that it only exists when \(P>m\). Moreover, it is linearly stable for \(S<m\) and marginally stable otherwise. 

\subsection{Coexistence states}
As said in the main text, the coexistence state is the only possible steady state when \(\pi^+,\pi^->0\). For \(\pi^+=\pi^-=0\) the solution \(\rho=\frac{1}{2}\) may appear as an instance of an absorbing state.

When \(\pi^+,\pi^->0\) for some values of \(\rho\), we need conditions \eqref{eq:coex1} and \eqref{eq:coex2}. In this case, \(\frac{d}{dt}\rho=\frac{1}{t_0}(1-2\rho)\) which has \(\rho=\frac{1}{2}\) as the only fixed point, which in turn is linearly stable.

\subsection{Absorbing states}

The absorbing states appear for a \(\rho\) satisfying \(\pi^+=\pi^-=0\), which may include \(\rho=0,\frac{1}{2},1\) as special cases, as already mentioned. The previous conditions are equivalent to Eqs.~\eqref{eq:absorb1}--\eqref{eq:absorb2} and give rise to the following regions of parameters where the absorbing states can be reached:
\begin{itemize}
\item[+] For \(R=m\), \(S>m\), \(P=m\), and \(T>m\) for all values of \(\rho\).
\item[+] Case I: \(R>m\) and \(P>m\). Here there are four possible regions:
  \begin{itemize}
  \item[-] For \(\tau \ge 1/N,\, \sigma \ge 1/N \), and all \(\rho\).
  \item[-] For \(\tau \ge 1/N,\, \sigma < 1/N\), and \( \rho >\rho_{-\sigma} \).
  \item[-] For \(\tau < 1/N,\, \sigma < 1/N \), \(\rho_{-\sigma}<\rho <1-\rho_{-\tau}\) when \(\tau \sigma \lesssim 1\). In the border case \(\tau \sigma \simeq 1\), there is only one possible value of $\rho$.
  \item[-]  For \(\tau < 1/N,\, \sigma \ge 1/N \), and \(\rho <1-\rho_{-\tau}\). 
  \end{itemize}
  We have introduced the following notation
  \begin{equation}
    \rho_x\equiv \frac{1-\frac{1}{N}}{1+x}
  \end{equation}
  and used the approximate symbols when \(N\to \infty\) to alleviate the mathematical writing. 
\item[+] Case II: \(R>m,\, P<m\) with two regions:
  \begin{itemize}
  \item[-] For \(\tau >- 1/N,\, \sigma \ge 1/N \) when \(\rho >1-\rho_\tau\).
  \item[-] For \(\tau >- 1/N,\, \sigma < 1/N\) when \( \rho >\max(1-\rho_\tau,\rho_{-\sigma}) \).
  \end{itemize}
  Case II' follows immediately from this one. 
\item[+] Case III: \(R<m,\, P<m\) with only one region:
  \begin{itemize}
  \item[-] for \(\tau > -1/N,\, \sigma > -1/N \) when \(1-\rho_\tau <\rho <\rho_\sigma\), which requires \(\tau \sigma \gtrsim 1\).
  \end{itemize}
\end{itemize}

\subsection{Discontinuous states}

Let us identify a discontinuous state with \(\rho^*\). As mentioned in the main text, it is characterized by the following condition, which involves the effective force \(F\):
\begin{eqnarray}
  \label{eq:cond1}
  && F(\rho^{*-})>0, \\
  \label{eq:cond2}
  && F(\rho^{*+})<0.
\end{eqnarray}
In case of any of the two inequalities where and equality, then \(\rho^*\) would be also an absorbing state as well. 

In order to identify the possible values \(\rho^*\in(0,1)\) verifying conditions \eqref{eq:cond1} and \eqref{eq:cond2}, it is convenient to distinguish between two possibilities: \(\rho^*\in(0,\frac12)\) and \(\rho^*\in(\frac12,1)\). After some algebra and taking \(N\to \infty\) for simplicity, we identify different regions where discontinuous states are present:
\begin{itemize}
\item[+] For \(-k_d\tau>1\), \(k_d\tau \sigma>k_c\), with \(\rho^*\simeq \frac{1}{1-k_d\tau}\).
\item[+] For \(-k_c\sigma>1\), \(k_c\tau \sigma>k_d\), with \(\rho^*\simeq 1-\frac{1}{1-k_c\sigma}\).
\end{itemize}

\section{Dependence of the games on the mood}
\label{appen:2}

In this Appendix, we study the effect of changing the aspiration or mood parameter \(m\) in the Stag Hunt, Harmony, and Snowdrift games. The Prisoner's dilemma was already discussed in the main text. 

\subsection{Stag Hunt}

For the Stag Hunt game, we choose as representative parameters \(P=0,\, R=1,\, S=-\frac{1}{2}\), and \(T=\frac{1}{2}\). At the mean-field level, according to the results in \cite{khalil2023deterministic}, in the final state of the system with rational choices agents either always defect or cooperate. The actual final state is determined by the specific values of the parameters and the initial condition. With a small degree of irrationality, the behavior is very similar. Only when most decisions are irrational, the system approaches coexistence. 

\begin{figure}[!h]
  \begin{center}
    \includegraphics[width=0.99\linewidth]{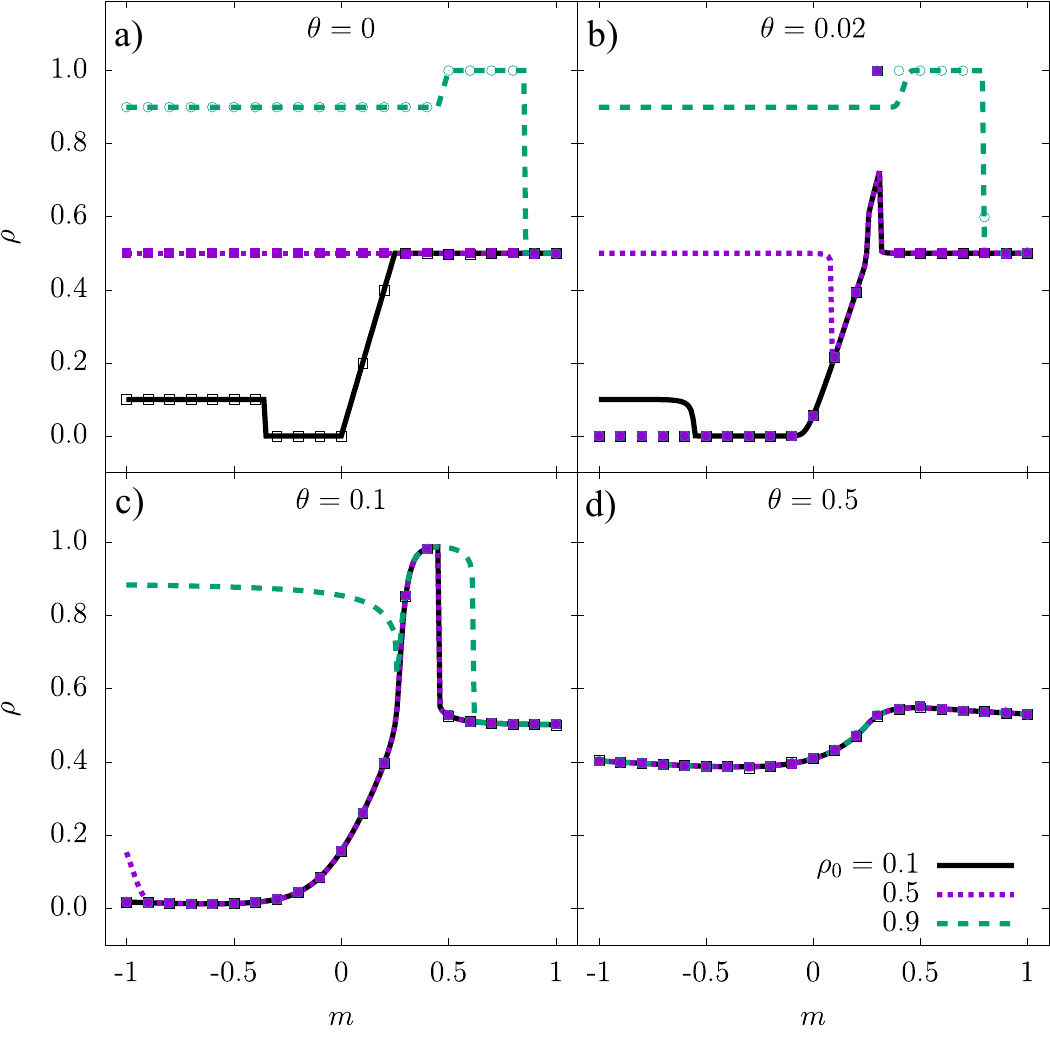}
  \end{center}
  \caption{Cooperation density \(\rho\) as a function of the mood \(m\) for the game parameters \(T=0.5;\, S=-0.5;\, P=0;\, R=1\) and the other parameters as in Fig.~\ref{fig:games1}. \label{fig:games2}}
\end{figure}

As for the Prisoner's dilemma, our results predict a much more complex behavior for the Stag Hunt model than the model in \cite{khalil2023deterministic}, as Fig.~\ref{fig:games2} clearly shows. Again, the final level of cooperation is strongly dependent on the temperature, the initial conditions, the evolution time, and specifically on the aspiration \(m\). Moreover, for this game, the system is more prone to suffer from abrupt changes. Interestingly, for a relatively high temperature \(\theta=0.1\) there is an optimal value of \(m\simeq 0.5\) that maximizes cooperation. 

\subsection{Harmony}

\begin{figure}[!h]
  \begin{center}
    \includegraphics[width=0.99\linewidth]{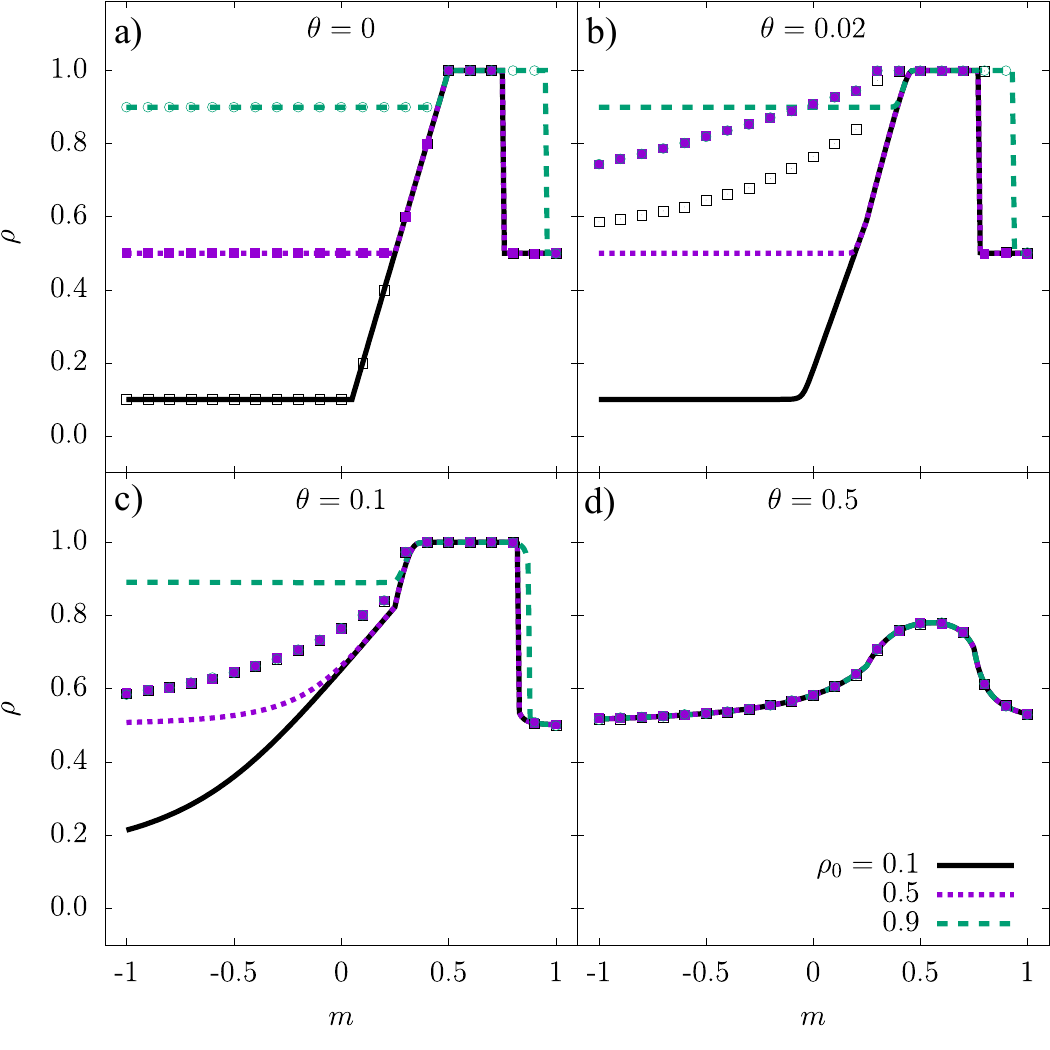}
  \end{center}
  \caption{Cooperation density \(\rho\) as a function of the mood \(m\) for the game parameters \(T=0.5;\, S=0.5;\, P=0;\, R=1\) and the other parameters as in Fig.~\ref{fig:games1}. \label{fig:games3}}
\end{figure}

The explicit dependence of the Harmony game on \(m\) for the specific game parameters \(P=0,\, R=1,\ S=T=\frac{1}{2}\) and different temperatures and initial conditions are given in Fig.~\ref{fig:games3}. As can be seen, the game shows rich behavior, with nontrivial dependence on the parameters. For zero effective temperature (rational choices), the cooperation density can take any value, provided the mood is small enough, as in the previous games. Here, however, we found an intermediate interval of \(m\) for which the system reaches complete cooperation. This optimal behavior remains with some degree of irrationality (positive effective temperature). This interesting behavior is in contrast with the usual one found when the dynamics is driven by comparing payoffs, as in \cite{khalil2023deterministic}, where cooperation consensus is the final state of the system for small and moderate degree of irrationality (say \(\theta\leq 0.5\)). 

\subsection{Snowdrift}

\begin{figure}[!h]
  \begin{center}
    \includegraphics[width=0.99\linewidth]{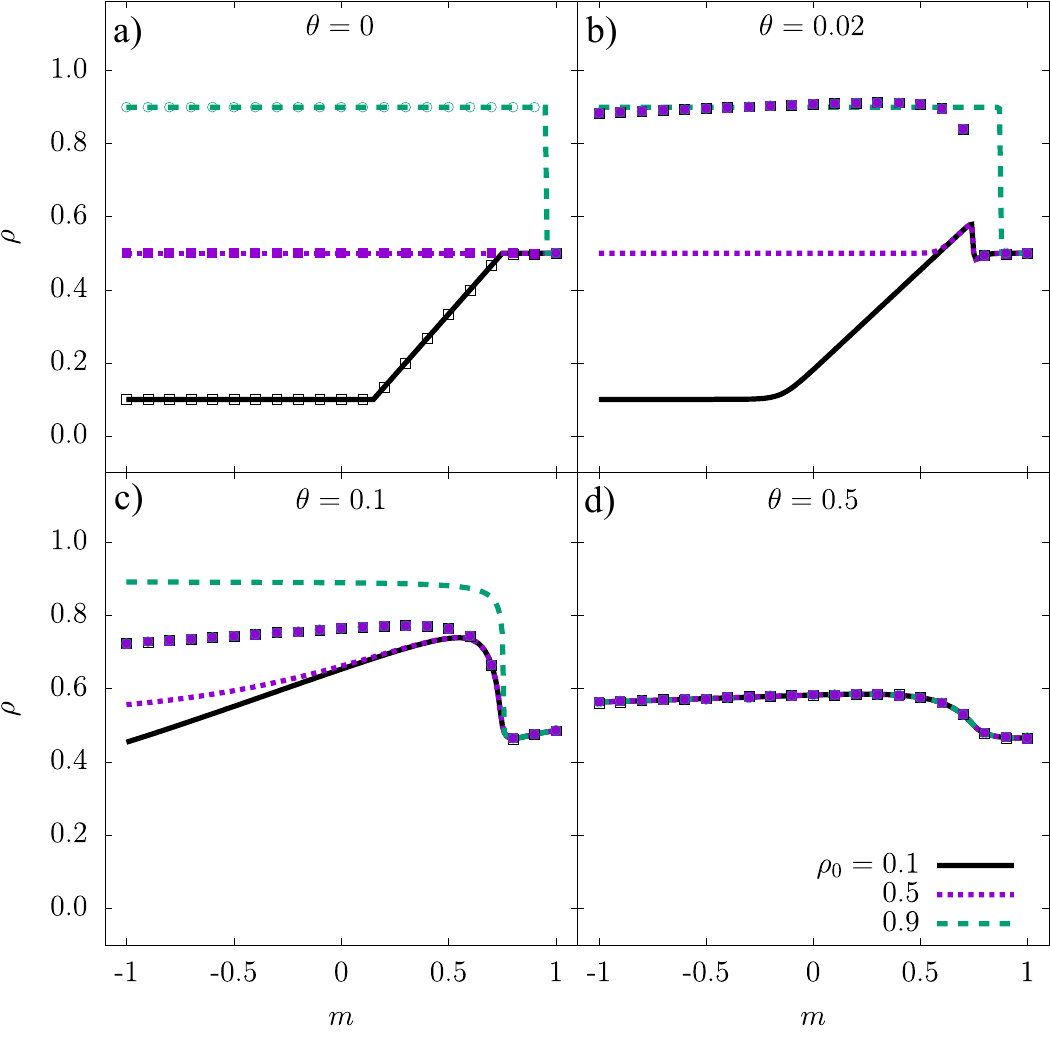}
  \end{center}
  \caption{Cooperation density \(\rho\) as a function of the mood \(m\) for the game parameters \(T=1.5;\, S=0.5;\, P=0;\, R=1\) and the other parameters as in Fig.~\ref{fig:games1}. \label{fig:games4}}
\end{figure}

Once again, in the Snowdrift game, we observe important differences between the results obtained in this work and that in Ref.~\cite{khalil2023deterministic}. Here, as it is apparent from \ref{fig:games4} for the specific parameters \(P=0,\, R=1,\, S=\frac{1}{2}\), and \(T=\frac{3}{2}\), the level of cooperation can take any value for rational behavior (small temperature) and low aspiration (small mood), while for larger effective temperatures the cooperation density is almost independent of the mood, except for a small range around \(m=\frac34\).

\clearpage

\bibliography{references}

\end{document}